\documentclass[journal]{IEEEtran}
\usepackage{cite}
\usepackage{amssymb,amsmath}
\usepackage{hyperref}
\usepackage{graphicx}
\usepackage{algorithm}
\usepackage{algorithmicx}
\usepackage{algpseudocode}
\usepackage{array, booktabs}
\usepackage{caption, subcaption}
\usepackage{color}
\newcommand{\red}[1]{\textcolor{black}{#1}}

\usepackage{setspace}
\usepackage{makecell}
\newcolumntype{Y}{>{\centering\arraybackslash}X}

\ifCLASSINFOpdf
\else
\fi

\begin{document}
%
\title{Learning the policy for mixed electric platoon control of automated and human-driven vehicles at signalized intersection: a random search approach}
%
%
%

\author{Xia Jiang,
        Jian Zhang,~\IEEEmembership{Member,~IEEE,}
        Xiaoyu Shi,
        and Jian Cheng
\thanks{Manuscript received December 28, 2021. (Corresponding author is Jian Zhang.)}
\thanks{X. Jiang and X Shi was with the School of Transportation, Southeast University, Nanjing, Jiangsu 210096, China (e-mail:  summer142857.jiang@gmail.com;2373497459@qq.com).}
\thanks{J. Zhang is with the School of Transportation, Southeast University, Nanjing, Jiangsu 210096, China and the School of Engineering, Tibet University, Lhasa, Tibet 850000, China (e-mail: jianzhang@seu.edu.cn)}
\thanks{J. Cheng is with Nanjing Les Information Technology Co.Ltd, Nanjing, Jiangsu 210096, China(e-mail: cheng\_j@les.cn)}
}

%
%

\markboth{IEEE Transactions}%
{Shell \MakeLowercase{\textit{et al.}}: Bare Demo of IEEEtran.cls for IEEE Journals}

\maketitle

\begin{abstract}
The upgrading and updating of vehicles have accelerated in the past decades. Out of the need for environmental friendliness and intelligence, electric vehicles (EVs) and connected and automated vehicles (CAVs) have become new components of transportation systems. This paper develops a reinforcement learning framework to implement adaptive control for an electric platoon composed of CAVs and human-driven vehicles (HDVs) at a signalized intersection. Firstly, a Markov Decision Process (MDP) model is proposed to describe the decision process of the mixed platoon. Novel state representation and reward function are designed for the model to consider the behavior of the whole platoon. Secondly, in order to deal with the delayed reward, an Augmented Random Search (ARS) algorithm is proposed. The control policy learned by the agent can guide the longitudinal motion of the CAV, which serves as the leader of the platoon. Finally, a series of simulations are carried out in simulation suite SUMO. Compared with several state-of-the-art (SOTA) reinforcement learning approaches, the proposed method can obtain a higher reward. Meanwhile, \red{the simulation results demonstrate the effectiveness of the delay reward, which is designed to outperform distributed reward mechanism.} Compared with \red{normal car-following behavior}, the sensitivity analysis reveals that the energy can be saved to different extends \red{(39.27\%-82.51\%)} by adjusting the relative importance of the optimization goal. \red{On the premise that travel delay is not sacrificed,} the proposed control method can save \red{up to 53.64\%} electric energy.

\end{abstract}


%
\IEEEpeerreviewmaketitle

\section{Introduction}
%
%
%
%
\IEEEPARstart{T}{he} advancements in artificial intelligence (AI), communication technologies, and vehicular technology have promoted the automation and electrification of vehicles in recent years. Automation nurtures the creation of connected and automated vehicles (CAVs), which is widely accepted as an effective way to improve traffic conditions \cite{FAGNANT2015167,7562449,7515222,ILGINGULER2014121,8758952}.  One problem associated with the application of CAVs to real world is that the design of the control strategy is uncertain, whereas the efficient functioning of the CAVs is based on their decision and control modules. The task is especially challenged in urban intersection scenarios, which are viewed as the bottlenecks of urban traffic, as they are the places where traffic flows with different directions converge. Since the operation of vehicles can be interrupted by traffic signals, the control law of CAVs at signalized intersections is crucial, \red{when it can determine the traffic performance in such urban scenarios.} Practical control approaches of CAVs \red{have shown that the travel efficiency, energy consumption, and safety can be improved at intersections \cite{8848852,GUANETTI201818, PENG2021100017}, so the significance of CAV-related research with regard to intersections is revealed.}

\red{In terms of research topic, a majority of studies focus on a traffic environment with 100\% CAV penetration rate, in which the conventional traffic signals can be eliminated, because the information of vehicular traffic can be completely obtained in real time \cite{OLOVSSON2022100056} and the vehicles can be controlled in a centralized manner \cite{8610389, YU2019416, 9611165}. Despite the fact that a pure CAV environment can create an unprecedented intelligent transportation system (ITS), there is a general consensus among researchers about the inevitability of the coexistence of CAVs and human-driven vehicles (HDVs) \cite{6907953, 9352231,ZHU2021126368,8965243,9345989,SHARMA2021102934}. Given this, controlling individual CAV at intersections becomes a promising way to exploit the potential advantages that CAVs can bring to urban transportation system. Having CAVs under control by embedded controller, relative indicators, such as travel time, energy consumption, and traffic safety, can be optimized for individual vehicle. \cite{8463636,mousa2020,bai2022hybrid}} 

\red{On the other side, the communication ability of CAVs makes it possible to implement cooperative control of several individual CAVs in a mixed traffic environment, which is usually achieved by platooning to extend the beneficial effect from vehicle level to platoon level \cite{8704319, MA2021102746,9352231,1692962,8917007,8735723}. The cooperative control approaches are capable of generating smoother trajectories and energy-saving speed profiles for CAVs. However, the application of automated driving system and vehicle-to-infrastructure (V2I) communication should not only enable the intelligent vehicles to make better decisions and enhance its own functionality \cite{zhang2021study}, but also improve the overall traffic performance, instead of sacrificing the mobility or energy consumption of other HDVs. Whereas the operation of CAVs may has a direct impact on other HDVs, and sometimes this influence would interfere with the normal running of those controlled by human drivers \cite{yue2021effects}, leveraging CAVs in mixed traffic condition to avert negative impact and promote the performance of HDVs is crucial, and this topic with mixed traffic is rarely discussed in urban intersection scenario. Zhao \emph{et al.} \cite{ZHAO2018802} proposed a framework that considers a mixed platoon of CAVs and HDVs at a signalized intersection to reduce the holistic energy consumption. Chen \emph{et al.} \cite{CHEN2021103138} explicitly made a definition of mixed platoon and formulated a control framework. }

\red{In addition to research topic, the formulation of control laws for CAV-related control problems is also important, usually obtaining by Model Predictive Control (MPC) \cite{5454336, GONG201825, ZHAO2018802} or Dynamic Programming (DP) \cite{5663859, 8550416}, which are challenged with computation complexity. Similarly, the aforementioned mixed platoon control framework are all based on a perspective of optimal control theory by expressly embodying cost functions, constraints, and solving algorithms.} It is also pointed out that these model-based methods need to simplify the dynamics of the environment or decompose the control problem into several sub-processes  \cite{GUO2021102980}. Accordingly, the lack of accuracy and generalization ability of the methods can impose an adverse impact on their practical application. To achieve cost-efficient in terms of computation time, some rule-based approaches are studied \cite{CI2019672,zhao2020dynamic,6083084}, but the optimality can not be ensured. With the intent to implement adaptive control with real-time ability, more competent approaches are supposed to be developed.

The Deep Reinforcement Learning (DRL) algorithms recently brought about new solutions for the vehicular control problem \cite{SHI2021103421}. Benefiting from the strong fitting ability of deep neural networks (DNNs), the DRL technique has the potential to approximate the optimal control process. In the DRL theory, an agent can choose actions according to the observed states so as to maximize its expected accumulated reward. For the general traffic control problems, the reward can be energy consumption, traffic delay, or the combination of relevant indicators. Based on the DRL algorithms, a few frameworks have been proposed in recent years to control CAVs at the proximity of signalized intersections. Shi \emph{et al.} \cite{application2018} applied Q-learning to improve the fuel consumption efficiency of a connected vehicle at a signalized intersection. An improved version of Q-learning based control framework, integrating with a deep Q network (DQN), was developed by Mousa \emph{et al.} \cite{mousa2020}. However, as one of the value-based DQL algorithms, the DQN approach cannot deal with the problems with continuous action space. Therefore, they directly took discrete velocity change rate as the action space, which can result in a local optimum solution. With the application of policy-based algorithms, the aforementioned problems can be tackled. Guo \emph{et al.} \cite{GUO2021102980} utilized a deep deterministic policy gradient (DDPG) algorithm to implement continuous longitudinal control of a CAV. Similarly, Zhou \emph{et al.} \cite{8848852} also trained DDPG agents to develop an efficient and energy-saving car following strategy. Furthermore, based on DDPG algorithm, they demonstrated that the method could improve travel efficiency by reducing the negative impact of traffic oscillations \cite{QU2020114030}. Wegner \emph{et al.} \cite{WEGENER2021102967} and \red{Zhang \emph{et al.} \cite{Jian2022} had} explored the energy-saving potential of electric CAV at urban signalized intersections by employing a twin-delayed deep deterministic policy gradient (TD3) agent, which is trained to control itself adaptively.

Nevertheless, there are some drawbacks that do exist among the aforementioned policy-based \red{DRL} approaches. Firstly, they all used stepwise reward signals to facilitate the learning process, and the policy learned by the agent in this situation cannot be equivalent to global optimum. For example, the framework put forward by Guo \emph{et al.} used stepwise travel distance to surrogate the total travel time of a CAV in one episode \cite{QU2020114030}, while the value of travel time can only be acquired after the CAV crosses the intersection. Although the agent can obtain reward signal in distributed form for each simulation step, the combination of the travel distance of all the steps is not tantamount to the total travel time. Intuitively, the agent may encounter a red light if it chooses the action in such a greedy way (i.e., aiming to maximize its stepwise travel time). Secondly, the previous DRL-based studies focus on the performance of a single CAV and ignore the integrated control of several vehicles. The CAVs can produce selfish policies \red{in an "ego-efficient" way}, which cannot \red{guarantee improved} performance of mixed platoons. Finally, it is known that algorithms like DDPG are highly sensitive to hyperparameter choices \cite{mania2018simple}. The traditional DRL approaches can also suffer from the sample efficiency problem, especially for the delayed reward situation. Therefore, a more effective method should be built to promote the application of reinforcement learning in this domain.

To address the above issues, this article develops a novel reinforcement learning control framework for CAVs at signalized intersections. A delayed reward Markov Decision Process (MDP) is formulated to describe the mathematical model of the \red{ control task in terms of the longitudinal motion of the platoon}. The state of the MDP considers the \red{leading} CAV and \red{its following} HDVs in a mixed platoon. \red{With regard to the reward signal, this paper define that it} can only be obtained when the platoon crosses the junction\red{, and simulation studies would manifest the benefits of the setting}. With the intent to deal with the delayed reward, an augmented random search (ARS) algorithm is proposed for the agent learning the control policy. The learning and evaluation of the framework are carried out in SUMO platform \cite{8569938}, which can demonstrate the effectiveness of the proposed method through microscopic traffic simulations. 

\red{Moreover, this paper takes the electric mixed platoons as research objects and make effort to optimize its electricity consumption. The starting point of electric vehicles is based on following reasons:} (1) The electrification of vehicles shows great promise to achieve sustainable traffic development \cite{FERRERO2016450}, as the carbon emissions and air pollution caused by the transportation system is still rising \cite{ZHANG2021116215}. (2) Due to the regenerative braking energy of electric vehicles (EVs), the control of electric CAVs is more challenging than traditional gasoline cars. At the same time, the EVs show a higher potential of energy conversion efficiency at low load range \cite{Hideki2016}. \red{In this case, the research of electric mixed platoon would have realistic meaning for a electric and intelligent road transportation system in the near future.}

The remainder of this paper is structured as follows. Section \uppercase\expandafter{\romannumeral2} introduces the preliminaries of DRL and the car-following model of HDVs. Section \uppercase\expandafter{\romannumeral3} provides the MDP formulation of the platoon-based control strategy. Section \uppercase\expandafter{\romannumeral4} proposes the ARS algorithm to implement the self-learning mechanism. Section \uppercase\expandafter{\romannumeral5} reports a series of simulations carried out in the SUMO software and makes a comparison study with several state-of-the-art (SOTA) methods. Finally, some concluding remarks are presented in Section \uppercase\expandafter{\romannumeral6}.

\section{Preliminary}
\subsection{Background of DRL}
Reinforcement learning is an important branch of machine learning. The object to be controlled in reinforcement learning is seen as an agent, and the learning process can be promoted by a series of agent-environment interactions. One complete play of the agent interacting with the environment is called as an episode. Generally, in step $t$ of an episode, an agent can observe a state $s_t$, which is usually a feedback by the environment. Then, the agent can conduct an action $a_t$ according to its policy $\pi(a_t|s_t)$. As a result, the agent can obtain a reward signal $r_t$, which is usually the representation of its optimization goal. Note that $r_t$ can be sparse when the reward can only be acquired in the terminal stage (i.e. with delayed rewards).

The process can be basically given by the MDP, which is defined as a five-tuple $(\mathbb{S},\mathbb{A},\mathbb{R},\mathbb{P},\gamma)$. $\mathbb{S}$, $\mathbb{A}$, and $\mathbb{R}$ denote the state space, the action space, and the reward space of the agent, respectively. For each timestep $t$, we have $s_t \in \mathbb{S}$, $a_t \in \mathbb{A}$, and $r_t \in \mathbb{R}$. Meanwhile, $\mathbb{P}$ specifies the state transition probability function: $\mathbb{S} \times \mathbb{S} \times \mathbb{A} \rightarrow [0,\infty)$, which can emit the probability density of the next state $s_{t+1} \in \mathbb{S}$ given the current state $s_t \in \mathbb{S}$ and action $a_t \in \mathbb{A}$. Moreover, $\gamma$ is a discount factor that measures the relative importance of the current reward and future reward. By interacting with the environment continuously, the agent aims to find an optimum policy that can maximize the expected sum of discounted future rewards $r^{\gamma}_{t}=r_{t+1}+\gamma r_{t+2}+\gamma^2r_{t+3}+...=\sum^{\infty}_{k=0}\gamma^k r_{t+k}$. For any policy $\pi$, the state-action value function is $Q^{\pi}(s,a)=\mathbf{E}[r^{\gamma}_{t}|s_t=s,a_t=a,\pi]$, where $a_{t+k} \sim \pi(\cdot|s_{t+k})$ for all $a_{t+k}$ and $s_{t+k}$ for $k \in [t+1,\infty)$. Meanwhile, the state value function is $v^{\pi}(s)=\mathbf{E}[r^{\gamma}_{t}|s_t=s,\pi]$. According to the Bellman equation, we have $v^{\pi}(s)=\sum_{a \in \mathbb{A}}\pi(a|s)Q^{\pi}(s,a)$. Finally, let $\Pi$ represent the set of all possible policies, and the optimal policy $\pi^*$ can be defined as:
\begin{equation} \label{eq:eq0}
    \pi^* \in \arg \max_{\pi \in \Pi}\mathbf{E}[r^{\gamma}_{t}|\pi]
\end{equation}
As a result, the agent can always select the optimal action following the optimal policy $\pi^*$.

The DRL technique makes use of deep learning to promote the traditional reinforcement learning approaches. Suppose the set of parameters of the utilized neural network is $\theta$, we can parameterize the state-action value function by $Q(s,a|\theta) \approx Q^*(s,a)$ for the value-based DRL algorithms, in order to approximate the optimal state-action value function $Q^*(s,a)$. As for the policy-based DRL algorithms, the policy is directly parameterized as $\pi(s,a|\theta)$. Hereafter, the learning process will adjust the set of parameters $\theta$ according to the "trial-and-error" mechanism to search for a suitable policy.

\subsection{Car Following Model of HDVs}
In this paper, we adopt the Intelligent Driver Model (IDM) to simulate the driving behavior of human drivers \cite{PhysRevE62}, whereas the model is widely used in microscopic traffic simulations \cite{8500556,SHARMA2019256,calvert2017will}. The acceleration of the $n_{th}$ vehicle at time $t$ is related to its current velocity, time headway, and the velocity of the front vehicle. The mathematical form of IDM is defined by \autoref{eq:eq1} and \autoref{eq:eq2}.

\begin{equation}\label{eq:eq1}
    a_n(t) = \frac{dv_n(t)}{dt} = a_0(1-(\frac{v_n(t)}{v_0})-(\frac{s^*_n(t)}{s_n(t)})^2)
\end{equation}

\begin{equation}\label{eq:eq2}
    s^*_n(t) = s_0 + Tv_n(t) + \frac{v_n(t)\Delta v(t)}{\sqrt{2a_0b}}
\end{equation}
where, $a_0$ and $v_0$ are the maximal acceleration and the expected velocity of the vehicle in free flow; $v_n(t)$ denotes the velocity of vehicle $n$ at time $t$; $s^*_n(t)$ and $s_n(t)$ are the expected headway and the real headway between the vehicle and its front vehicle, respectively; $s_0$ represents the minimal headway; $T$ denotes the safe time headway; $\Delta v(t)$ denotes the velocity difference between the vehicle and its leading vehicle. Finally, $b$ denotes an acceptable comfort-related deceleration.

\section{Markov Decision Process for the Problem}
\subsection{Problem Description}
\begin{figure}[htb]
    \centering
    \includegraphics[width=0.5\textwidth]{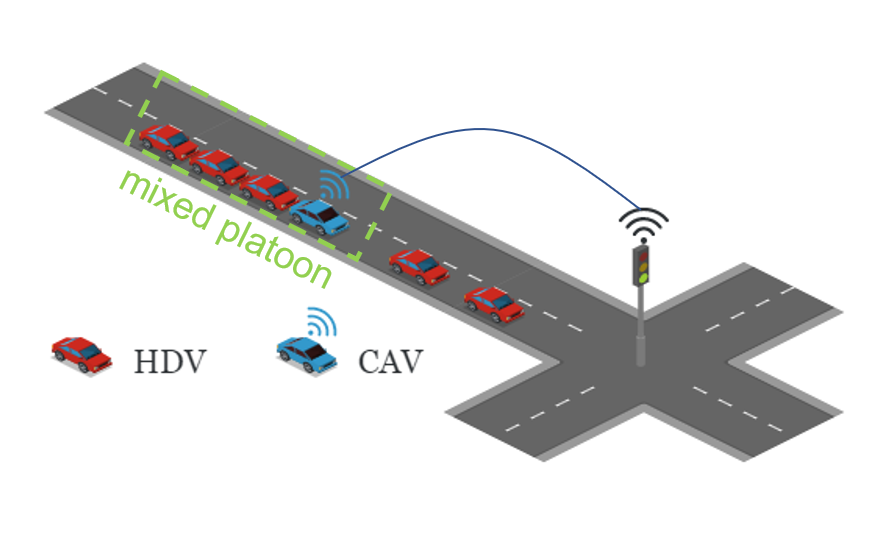}
    \caption{The illustration of the studied scenario.}
    \label{fig:fig1}
\end{figure}
As shown in \autoref{fig:fig1}, this study mainly focuses on a "1+n" form of the mixed platoon, consisting of one leading CAV and $n$ following HDVs. We call the electric CAV of the platoon "ego CAV", while the platoon led by the ego CAV is called "ego platoon". Besides the ego platoon, there are some other HDVs travel on the road, and this can make the simulation get close to the real traffic situation. In order to simplify the problem without losing any generality, we make some assumptions as below:

1) With the support of V2I communication, the ego CAV can obtain the Signal Phase and Timing (SPaT) information of the fixed-timing traffic signal.

2) The ego CAV can get the position, velocity, and acceleration of itself by vehicular operation system.

3) The positions and velocities of the HDVs belonging to the ego platoon can be obtained by the ego CAV. At the same time, the ego CAV can also get these data of its leading vehicle if the position of the leading vehicle is in a predefined range. The assumption can usually be achieved by the vehicle-to-vehicle (V2V) communication, \red{roadside units}, or the perception ability of the CAV \cite{ZHAO2018802, CHEN2021103138}.

Since the operation of the mixed platoon can be interrupted by other HDVs or traffic signals, the goal of the platoon is to reduce the overall delay and electric energy consumption. We basically study the longitudinal motion of the vehicles\red{, because the unexpected lane changing may interfere with normal operations of other HDVs, especially in the vicinity of signalized intersections. Although the scenario presented in \autoref{fig:fig1} is a single-lane environment, the proposed framework can be conducted for CAVs in a decentralized fashion for multi-lane scenarios}. Accordingly, an effective control law will generate a speed profile for the leading CAV and consider the motion of the subsequent HDVs. In this case, unnecessary stops and oscillations can be avoided to achieve the energy-saving goal.

\subsection{Specification of the MDP}
The elements in the MDP model, including $S$, $A$, and $R$ should be specified to apply the DRL framework. For a "1+n" mixed platoon, the formation of the three factors can be defined as follows.

\textit{1) State:} State is the description of the agent in current situation. All of the vehicles within the mixed platoon should be taken into account as part of the state. Meanwhile, the potential leading vehicle can be taken into account, as the ego CAV should keep a safe gap and estimate the traffic ahead. With the intent to reduce unnecessary stop-and-go operation, the agent also needs the SPaT information of the first downstream traffic signal. Therefore, let $s_t^C$, $s_t^H$, $s_t^L$, and $s_t^S$ be the CAV-related part, the HDVs-related part, the leading-vehicle-related part, and the signal-related part of the state, the state can be expressed as:
\begin{equation}\label{eq:eq3}
    s_t=(s_t^C, s_t^H, s_t^L, s_t^S)^T
\end{equation}

In this case, the details of each part of the state are shown as follows:
\begin{equation}\label{eq:eq4}
    s_t^C = (d(t), v(t))
\end{equation}
\begin{equation}\label{eq:eq5}
    s_t^H = (x_1(t), v_1(t), x_2(t), v_2(t),\dots, x_n(t), v_n(t))
\end{equation}
\begin{equation}\label{eq:eq6}
    s_t^H = (\Delta x(t), \Delta v(t), \Delta a(t))
\end{equation}
\begin{equation}\label{eq:eq7}
    s_t^H = (RT(t),E_s(t))
\end{equation}
where, $d(t)$ is the distance between the ego CAV and the stop line of the first downstream intersection at time $t$; $v(t)$ is the velocity of the ego CAV at time $t$; $x_i(t)$ and $v_i(t)$ are the lane position and speed of HDV $i$ for $i$ from $1$ to $n$. For the third item in \autoref{eq:eq3}, we set a predefined threshold $\chi_x$ to judge if there is a leading vehicle in front of the platoon. Let $L$ be the index of the potential leading vehicle, we set a boolean variable $\delta$ to identify the existence of the potential preceding vehicle:

\begin{equation}\label{eq:eq8}
    \delta = \begin{cases}
    True, & \text{if } x_L(t)-x(t) \leq \chi_x; \\
    False, & \text{otherwise.}
    \end{cases}
\end{equation}

The calculation of $\Delta x$, $\Delta v$, and $\Delta a$ according to the value $\delta$ are expressed as:
\begin{equation}\label{eq:eq9}
    \Delta x = \begin{cases}
    x_L(t)-x(t), & \text{if } \delta; \\
    \chi_x, & \text{otherwise.}
    \end{cases}
\end{equation}

\begin{equation}\label{eq:eq10}
    \Delta v = \begin{cases}
    v_L(t)-v(t), & \text{if } \delta; \\
    \chi_v, & \text{otherwise.}
    \end{cases}
\end{equation}

\begin{equation}\label{eq:eq11}
    \Delta a = \begin{cases}
    a_L(t)-a(t), & \text{if } \delta; \\
    \chi_a, & \text{otherwise.}
    \end{cases}
\end{equation}
where, $\chi_v$ and $\chi_a$ are the predefined default value of the two variables. In this paper, $\chi_x$ is set to $500m$, which means that the vehicle 500 meters away from the ego CAV will not affect its driving. Moreover, $\chi_v$ and $\chi_a$ are set to $13.88m/s$ and $7.5m/s^2$.

\begin{figure}[htb]
    \centering
    \includegraphics[width=0.35\textwidth]{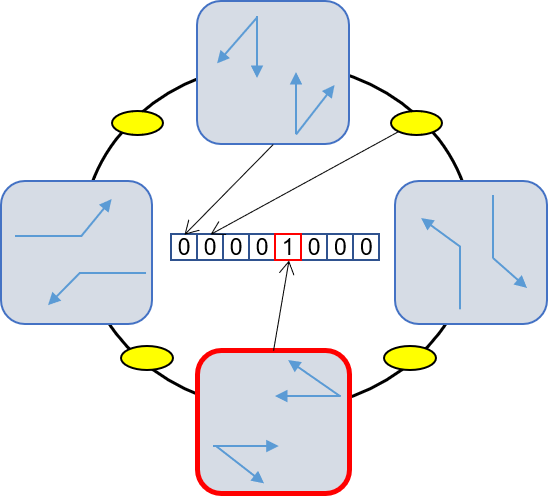}
    \caption{The phase diagram and its one-hot encoding.}
    \label{fig:fig2}
\end{figure}

As for the signal-related state, $RT(t)$ in \autoref{eq:eq7} denotes the remaining time of the current phase for the first downstream traffic signal, and this value can be retrieved in a communication environment. Furthermore, $E_s(t)$ denotes the one-hot encoding of the current phase of the traffic signal. The encoding process is illustrated in \autoref{fig:fig2}. The phase diagram shows the signal phase used in this study, and yellow light is added between two adjacent phases. If one phase is activated by the traffic signal (i.e., the phase with red box in \autoref{fig:fig2}), the corresponding element in the encoding vector will be set to 1, while other elements are all set to 0.

\textit{2) Action:} Due to the maneuverability of the system, the action is to change the acceleration of the ego CAV. Hence, the action space is constrained by the dynamics of the vehicle: $a_t \in [a_{min}, a_{max}]$, where $a_{min}$ and $ a_{max}$ are the maximal deceleration and acceleration of the vehicle. However, it is problematic to take the acceleration as the action directly. On the one hand, irrational accelerations will lead to unsafe operations like rear-end accidents, and this kind of phenomenon can occur very often during the training process; on the other hand, the speed of the vehicle may exceed the road speed limit with the effect of the action. Consequently, the modified action is stipulated as:
\begin{equation}\label{eq:eq12}
    a_t = \min(\tilde{a_t}, a^{IDM}(t))
\end{equation}
where, $\tilde{a_t}$ denotes the original acceleration value output by the DRL algorithm; $a^{IDM}(t)$ is the acceleration calculated by IDM. \autoref{eq:eq12} makes the acceleration of the ego CAV be kept in a safe range.

The velocity change of the ego CAV is defined as below to meet the speed limit $V_{max}$:
\begin{equation}\label{eq:eq13}
    v(t) = \max(\min(V_{max}, v(t-1)+a_t), 0)
\end{equation}
where, $v(t-1)$ is the speed of the ego CAV in last timestep.

\textit{3) Reward:} The optimization goal, including total energy consumption and travel delay, can only be calculated when the vehicles have crossed the signalized intersection. Distributing the delayed reward to each step in an episode is known as the temporal Credit Assignment Problem (CAP) \cite{4066245}, which is hard to deal with. The previous \red{studies} took stepwise energy consumption and travel distance to serve as a distributed proxy of the two parts of the delayed reward \red{\cite{GUO2021102980, Jian2022}}. Nevertheless, the cumulative travel distance cannot indicate the delay of the vehicles accurately. A more intuitive way is using the delayed reward, which can directly reflect the optimization goal. In this case, the reward is non-Markovian. In this study, we will show that our algorithm can commendably solve the CAP and train the agent. The reward function is defined as:
\begin{equation}\label{eq:eq14}
    r_t = \begin{cases}
        \sum_{i=0}^{n}-\omega_1 e_i - \omega_2 d_i, & \text{if } t=t_{final} \\
        0, & \text{otherwise.}
    \end{cases}
\end{equation}
where, $e_i$ denotes the total energy consumption of vehicle $i$; $d_i$ denotes the delay of vehicle $i$. Note that the vehicle with $i=0$ here represents the ego CAV. Meanwhile, $\omega_1$ and $\omega_2$ are weighting parameters that measure the relative importance of mobility indicator and energy indicator. Finally, $t_{final}$ specifies the finale of an episode. It is the time when the last HDV in the ego platoon crosses the intersection.

In \autoref{eq:eq14}, $e_i$ is calculated by a series of records in the whole episode. This study utilizes an energy model with energy brake-recovery mechanism embedded in SUMO to calculate the instantaneous electric consumption \cite{SUMO1007}. Note that any other energy model can be used owing to the generality of the proposed framework, even if a simple indicator that derived from the difference of the battery. The instantaneous energy consumption is calculated for each vehicle within the platoon in each step. Finally, $e_i$ is calculated when vehicle $i$ enter the intersection. Similarly, $d_i$ is expressed as:
\begin{equation}\label{eq:eq15}
    d_i = t_{f}^i - t_0 - \frac{L}{V_{max}}
\end{equation}
where, $t_f$ is the time when vehicle $i$ crosses the stop line of the junction;$t_0$ is the initial time; $L$ denotes the length of the entrance lane where the platoon locates.

\section{Augmented Random Search}
The purpose of the algorithm is to directly search a policy in continuous action space, while the obtained policy can approximate the optimal policy $\pi^*$ in \autoref{eq:eq0}. As the transition dynamics is unknown in most cases, model-free reinforcement learning algorithms are usually deployed. It is pointed out that many model-free DRL methods need too much data to search a proper optimization direction, and they can be very complicated without robustness \cite{mania2018simple}. Considering the practicability, we develop a ARS algorithm in this paper to search the policy in a black-box way. Being compared with the gradient-based DRL methods, the black-box optimization approach can achieve sample efficiency and have an advantage in cases with long action sequences and delayed reward \cite{salimans2017evolution}.

In the context of DRL, the policy is usually parameterized by a set of parameters $\theta$, which is supposed to be trained in training process. The ARS utilized a linear policy with parameter set $\theta$ instead of DNNs like most DRL algorithms. Note that throughout the rest of the paper we use $\pi_{\theta}$ to denote the ARS-based policy with parameter set $\theta$. Let the dimension of the state in \autoref{eq:eq3} be $p$. The parameter set $\theta$ is a $p \times n$ matrix, while the dimension of action is represented as $n$.

The update increment $\Delta \theta$ of $\theta$ follows:
\begin{equation}\label{eq:eq16}
    \Delta \theta = \frac{r(\pi_{\theta+\upsilon\mu},\xi_1) - r(\pi_{\theta-\upsilon\mu},\xi_2)}{\upsilon}
\end{equation}
where, $\xi_1$ and $\xi_2$ are random variables that encode the randomness of the environment; $\upsilon$ is a positive real number that denotes the standard deviation of the exploration noise; $\mu$ denotes a vector with zero mean Gaussian distribution. 

\begin{figure}[htb]
    \centering
    \includegraphics[width=0.4\textwidth]{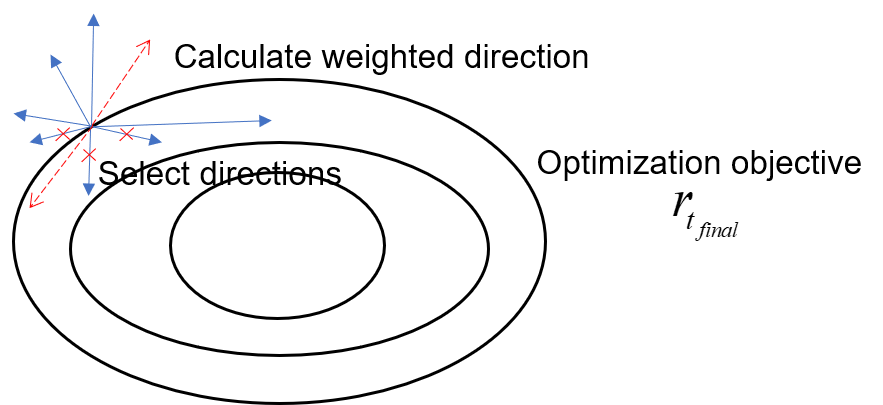}
    \caption{The sketch for the idea of ARS.}
    \label{fig:fig3}
\end{figure}

The basic idea of ARS is to randomly adds some tiny variables to the parameter $\theta$ along with the negative value of the corresponding value. After the perturbation, the variables with a higher reward have a bigger influence on the adjustment of $\theta$. This process is shown in \autoref{fig:fig3}. The directions with red crosses represent the variables with relatively low rewards, so they are eliminated when calculating the final updating direction. In particular, the red dashes represent the update direction weighted by the rest of the variables.

More specifically, The pseudocode of the proposed ARS is shown in \textbf{Algorithm 1}. Three tricks are adopted in the ARS algorithm to enhance its performance \cite{mania2018simple}:

\textit{1) Scaling by the standard deviation:} During the training process across the iterations, there will exist a large variation in the collected rewards record. In particular, the circumstance brings about difficulties for choosing a proper step size $\alpha$. In each iteration, $2K$ rewards are recorded. A standard deviation $\epsilon_R$ will be calculated and is used to scale the update step (see line 5 in \textbf{Algorithm 1}).

\textit{2) States normalization:} The purpose of normalization is to eliminate the influence of dimensional inconsistency of different elements in state vectors. For the parametric linear policy, it can promote non-isotropic explorations in the parameter space. For a perturbation direction $\mu$, there is:
\begin{equation}\label{eq:eq17}
    (\theta+\upsilon\mu)\text{diag}(\Sigma)^{-\frac{1}{2}}(x-\sigma)=(\acute{\theta}+\upsilon \mu \text{diag}(\Sigma)^{-\frac{1}{2}})(x-\sigma)
\end{equation}
where, $\acute{\theta}=\theta \text{diag}(\Sigma)^{-\frac{1}{2}}$

\textit{3) Using top-performing directions:} The perturbation direction $\mu$ is weighted by the difference of two opposed rewards $r(\pi_j,k,+)$ and $r(\pi_j,k,-)$ (see line 3 in \textbf{Algorithm 1}). Without this trick, the update steps push $\theta$ in the direction of $\mu_k$. However, using top-performing directions can order decreasingly the directions $\mu_k$ by $\max\{r(\pi_j,k,+), r(\pi_j,k,-)\}$. Finally, only the top $b$ directions are utilized to update the policy parameters (see line 5 in \textbf{Algorithm 1}).

During the iterations of training, only the total reward of an episode is used to evaluate the performance of a series of actions, so ARS can deal with maximally sparse and delayed rewards and avoid the difficulties produced by CAP. The feature makes it suitable to solve the platoon control problems with delayed reward configurations. Without the training of DNN, ARS can save much inference time, and it is promising to deploy such a computation-efficiency framework in real world.

\begin{algorithm}[t]\label{alg:algo1}
\caption{ARS for Mixed Platoon Control}
\hspace*{0.02in}{\bf Hyperparameters:}
step-size $\alpha$, number of directions sampled per iteration $K$, noise $\upsilon$, number of top-performing directions to use $b (b<K)$ \newline
\hspace*{0.02in}\bf{Initialize}:
$\theta_0=\mathbf{0} \in \mathbf{R}^{p \times n}$, $\sigma_0=\mathbf{0} \in \mathbf{R}$, $\Sigma_0=\mathbf{I_n} \in \mathbf{R}$, $j=0$
\begin{algorithmic}[1]
\While{end condition not satisfied}
    \State Sample $\mu_1, \mu_2,\dots,\mu_K$ in $\mathbf{R}^{p \times n}$ with i.i.d standard normal entries.
    \State Collect $2K$ episodes of horizon $H$ and their corresponding rewards using the $2k$ policies in SUMO:
    \Statex \qquad $\begin{cases}
        \pi_{j,k,+}(x) = (\theta_j+\upsilon\mu_k)\text{diag}(\Sigma_j)^{-\frac{1}{2}}(x-\sigma_j) \\
        \pi_{j,k,-}(x) = (\theta_j-\upsilon\mu_k)\text{diag}( \Sigma_j)^{-\frac{1}{2}}(x-\sigma_j) \\
    \end{cases}    
    $ \newline
    \Statex \qquad for $k \in {1, 2, \dots, K}$
    \State Sort the directions $\mu_k$ according to $\max{r(\pi_{j,k,+}, \pi_{j,k,-})}$. Let $\mu_{(k)}$ be the $k-$th largest direction, and by $\pi_{j,(k),+}$, $\pi_{j,(k),-}$ the corresponding policies.
    \State Update $\theta$ by step ($\epsilon_R$ denotes the standard deviation of the $2b$ rewards): \newline
    $\theta_{j+1} = \theta_j+\frac{\alpha}{b\epsilon_R}\sum_{k=1}^b [r(\pi_{j,k,+} - \pi_{j,k,-}]\mu_{(k)}$ 
    \State Set $\sigma_{j+1},\Sigma_{j+1}$ to be the mean and covariance value of the $2KH(j+1)$ states encountered from the start of training.
    \State$j \leftarrow j+1$
\EndWhile
\end{algorithmic}
\end{algorithm}
\section{Simulation Analysis}
As one of the most popular open-source traffic simulator, SUMO allows the modelling of the microscopic behavior of vehicles and pedestrians. The value of simulation in SUMO can be retrieved and changed through the "TraCI" interface by other program languages. In this study, a signalized intersection is built in SUMO environment, and the scenario is similar to that shown in \autoref{fig:fig1}.
\subsection{Simulation Settings}
The signal phases are shown in \autoref{fig:fig2}. As a fixed-timing traffic signal, the last time for each phase is $30s$, while a $3s$ yellow phase is inserted for every phase changing. Under the premise of comprehensive consideration of reality and generality, the other related parameters for simulation configuration are presented in \autoref{tab:table1}. Before the learning process of each episode, a pre-loading procedure is carried out. More precisely, the traffic volume is loaded for time $t_p$ (sampled from a uniform distribution) before the ego platoon enters the road, aiming at generating more dynamic traffic scenarios. Meanwhile, after a series of simulations, the hyperparameters of ARS are tuned manually. The standard deviation of parameter noise $\upsilon$ is set to 0.2; the number of directions sampled per iteration is set to 32. Note that the weighting parameters are set as: $\omega_1=6$, $\omega_2=1$ if no special explanation is provided. The sensitivity analysis of the two parameters are presented in the subsequent subsection.

\begin{table}[htb]
    \caption{The parameter setting of simulations}
    \label{tab:table1}
    \centering
    \resizebox{0.5\textwidth}{!}{
    \begin{tabular}{ccc}
    \hline\hline
    Item & Value & Unit \\
    \midrule
    Maximum acceleration of vehicles $a_{max}$ & 3.0 & $m/s^2$ \\
    Minimum acceleration of vehicles $a_{min}$ & -4.5 & $m/s^2$ \\
    Road speed limit $V_{max}$ & 13.88 & $m/s$ \\
    The length of the lane $L$ & 500 & $m$ \\
    Safe time headway $T$ in IDM & 1 & $s$ \\
    Acceptable comfort-related deceleration $b$ in IDM& -2.8& $m/s^2$ \\
    Hourly traffic volume & 400 & $veh$ \\
    Pre-loading time $t_p$ & $\mathcal{U}(180,220)$ & $s$ \\
    \hline\hline
    \end{tabular}
    }
\end{table}

\subsection{Training Performance}

\begin{figure}[htb]
    \includegraphics[width=0.48\textwidth]{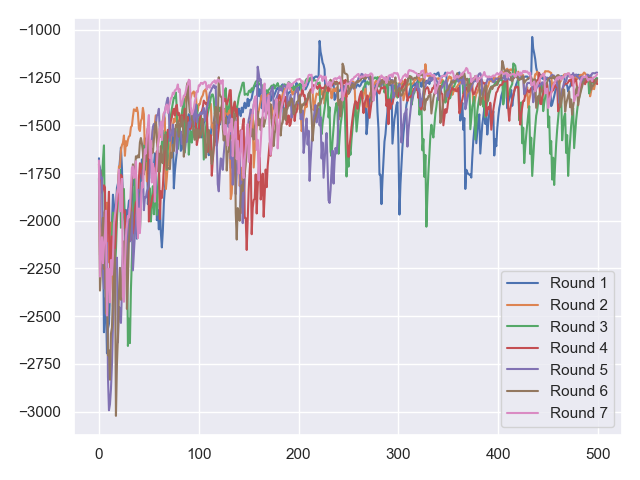}
    \caption{The smoothed episode rewards from seven rounds of training.}
    \label{fig:fig4}
\end{figure}

Firstly, with the intent to show the robustness of the ARS approach, we conduct 7 rounds of independent training with different random seeds. \autoref{fig:fig4} illustrates the training results. Owing to the noise rewards for different episode, a moving average function is applied to smooth the tendency: $R_k \leftarrow 0.8R_{k-1} + 0.2R_k$, where $k$ denotes the $k$-th episode. Although the fluctuation range of each round can be different, they can all converge to the same result with about -1250 reward.

\begin{figure}[htb]
    \includegraphics[width=0.48\textwidth]{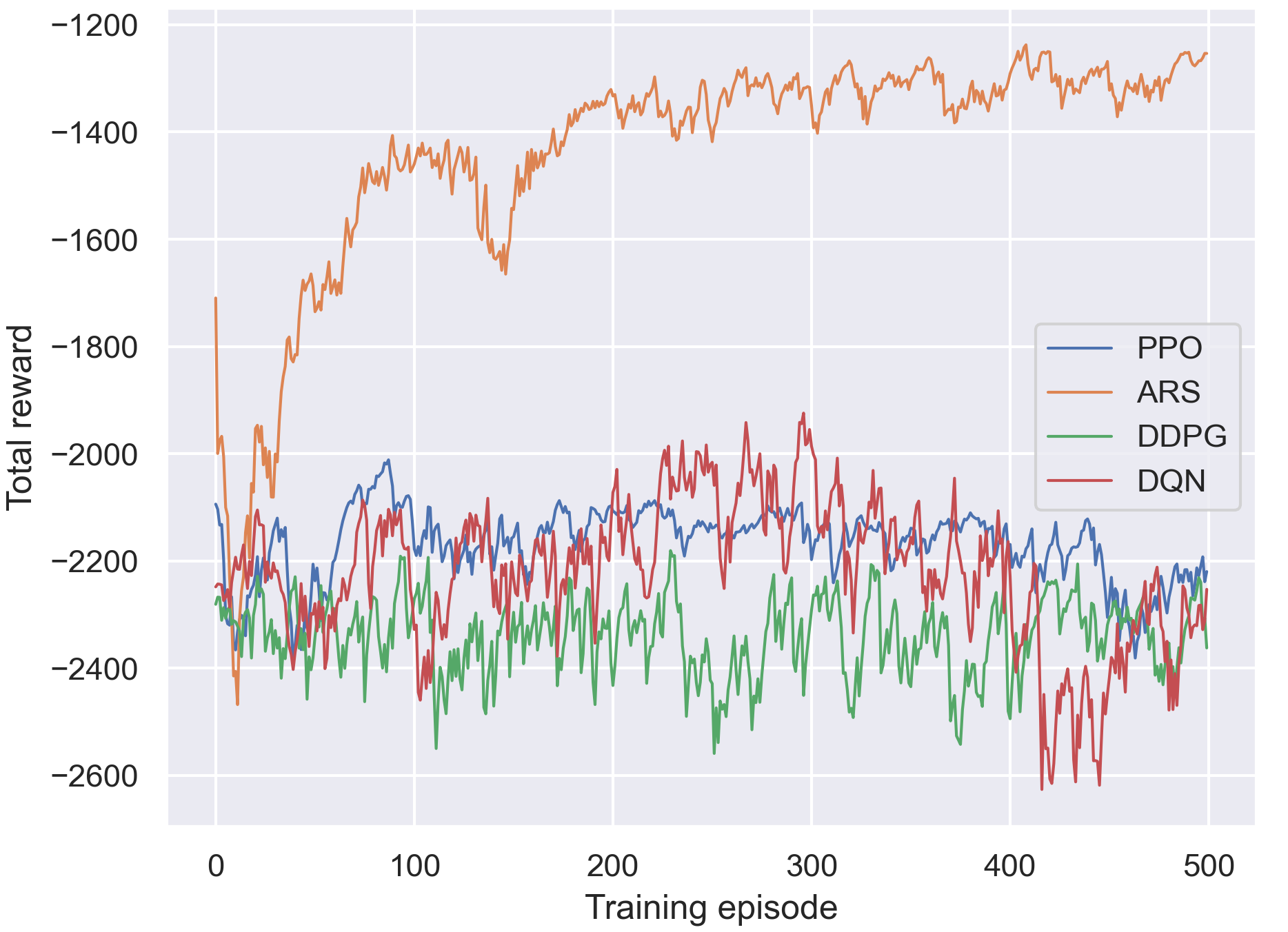}
    \caption{The comparison for different algorithms of the training process.}
    \label{fig:fig5}
\end{figure}

Secondly, we make a comparison study with other SOTA methods, including Proximal Policy Optimization (PPO), DDPG, and DQN. For each algorithm, the hyperparameters are tuned manually through several simulations, and the training results from seven independent are aggregated to obtain the final result to reduce the effect of randomness. Note that the action space of the DQN is set to a 16-length vector, which varies from $a_{min}$ to $a_{max}$ with the step of $0.5m/s^2$.  Taking the scenario with "1+3" mixed platoon as an example, the training processes are shown in \autoref{fig:fig5}. It can be seen that it is hard to train the agent for the other three SOTA algorithms with the delayed reward cases. However, the reward of the ARS agent can converge to a higher value compared with the other approaches.

\subsection{Exploring the Impact of Reward Configuration}

\begin{figure}[htb]
    \includegraphics[width=0.48\textwidth]{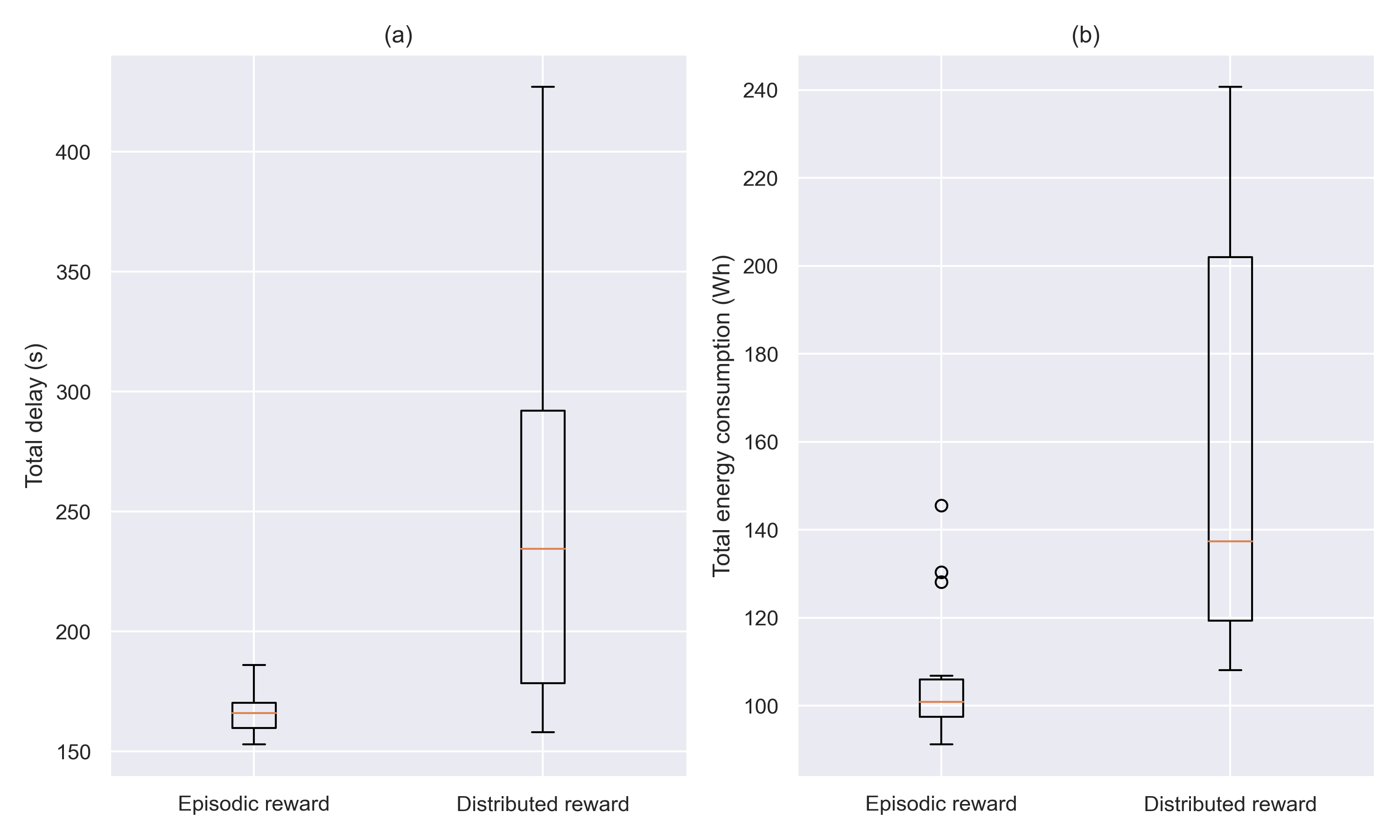}
    \caption{The comparison for different reward settings. (a) Total delay. (b) Total energy consumption.}
    \label{fig:fig6}
\end{figure}

To investigate the influence of reward settings, we compare the cases with episodic reward (ER) and distributed reward (DR) settings. In DR setting, the reward is calculated step by step according to the stepwise sum of energy consumption and travel distance of the platoon. Accordingly, we train the ARS agent five times, and collect five episodes of reward for each trained agent. More precisely, 25 groups of simulations are carried out to record the data for each reward setting. The results are presented in \autoref{fig:fig6}. Whether in terms of travel delay or electric consumption, the ER setting can outperform the DR setting. The agents with DR show high variance with respect of energy consumption indicator, and this illuminates the instability of this kind of configurations.

Similar studies can be conducted for PPO algorithm. \autoref{tab:table2} shows the mean value of the indicators, deriving from 25 episodes of simulations. IDM is introduced to serve as a baseline. In this case, the ego CAV is controlled by the IDM, which can represent the general car following scenario.

\begin{table}[tbp]
    \caption{Numerical results for different algorithm and reward settings}
    \label{tab:table2}
    \centering
    \resizebox{0.5\textwidth}{!}{
    \begin{tabular}{ccc}
    \hline\hline
    Method & Delay per Vehicle (s) & Total Electricity (Wh) \\
    \midrule
    PPO with DR & 143.03 & 222.77 \\
    PPO with ER & 241.58 & 246.55 \\
    ARS with DR  & 239.78 & 156.85 \\
    ARS with ER & 165.22 & \textbf{104.81} \\
    IDM & \textbf{57.98} & 612.42 \\
    \hline\hline
    \end{tabular}
    }
\end{table}

\autoref{tab:table2} demonstrates that the ER-based ARS can reduce energy consumption to the maximum extent. The DR-based PPO has a similar performance with the ER-based ARS in terms of total delay. However, ARS can reduce the electric energy consumption by 52.95\% compared with the DR-based PPO for a "1+3" platoon on average. Inevitably, the optimization on energy will lead to the sacrifice of mobility \cite{GUO2021102980}. With the setting of $\omega_1=6$ and $\omega_2=1$, 82.89\% energy is saved by the adaptive control implemented by ARS algorithm compared with IDM. The agent in this case behaves toward an extreme energy efficiency direction. Nevertheless, we will conduct a sensitivity analysis in the following subsection. The analysis can reveal that the agent can reduce energy consumption with almost no sacrifice of mobility.

\subsection{Performance for Different Platoon Size}
\begin{figure}[htb]
    \includegraphics[width=0.48\textwidth]{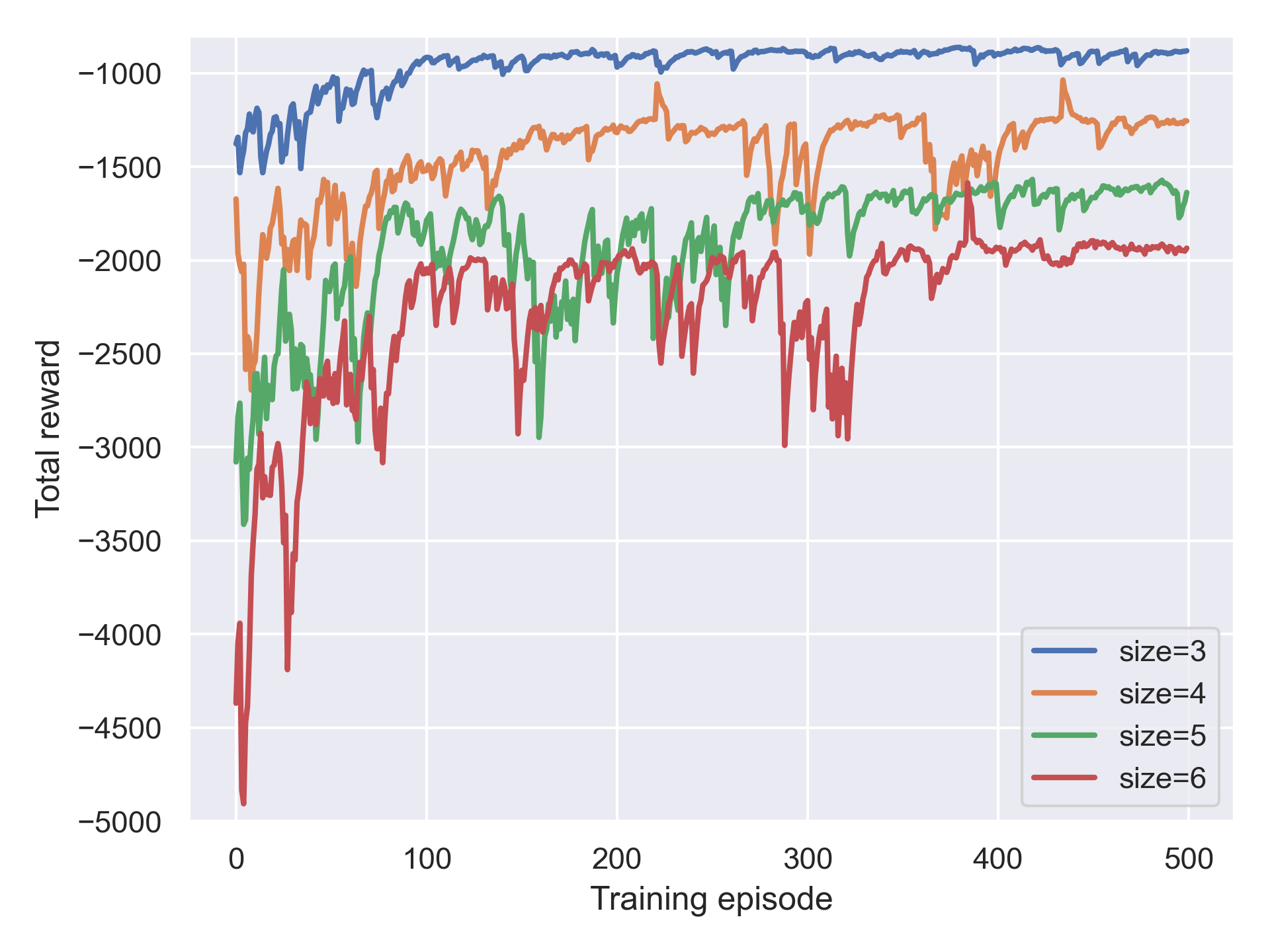}
    \caption{The training curve of the cases with different platoon size.}
    \label{fig:fig7}
\end{figure}

\autoref{fig:fig7} shows the smoothed training curve for the scenarios with different platoon sizes. It can be concluded that the cases with different platoon sizes can be optimized, and the results can converge to different values. The more HDVs are considered by the agent, the higher optimization rate can be observed. As a result, the framework has the potential to normally extend to multi-vehicle systems.

\begin{figure}[htb]
    \includegraphics[width=0.5\textwidth]{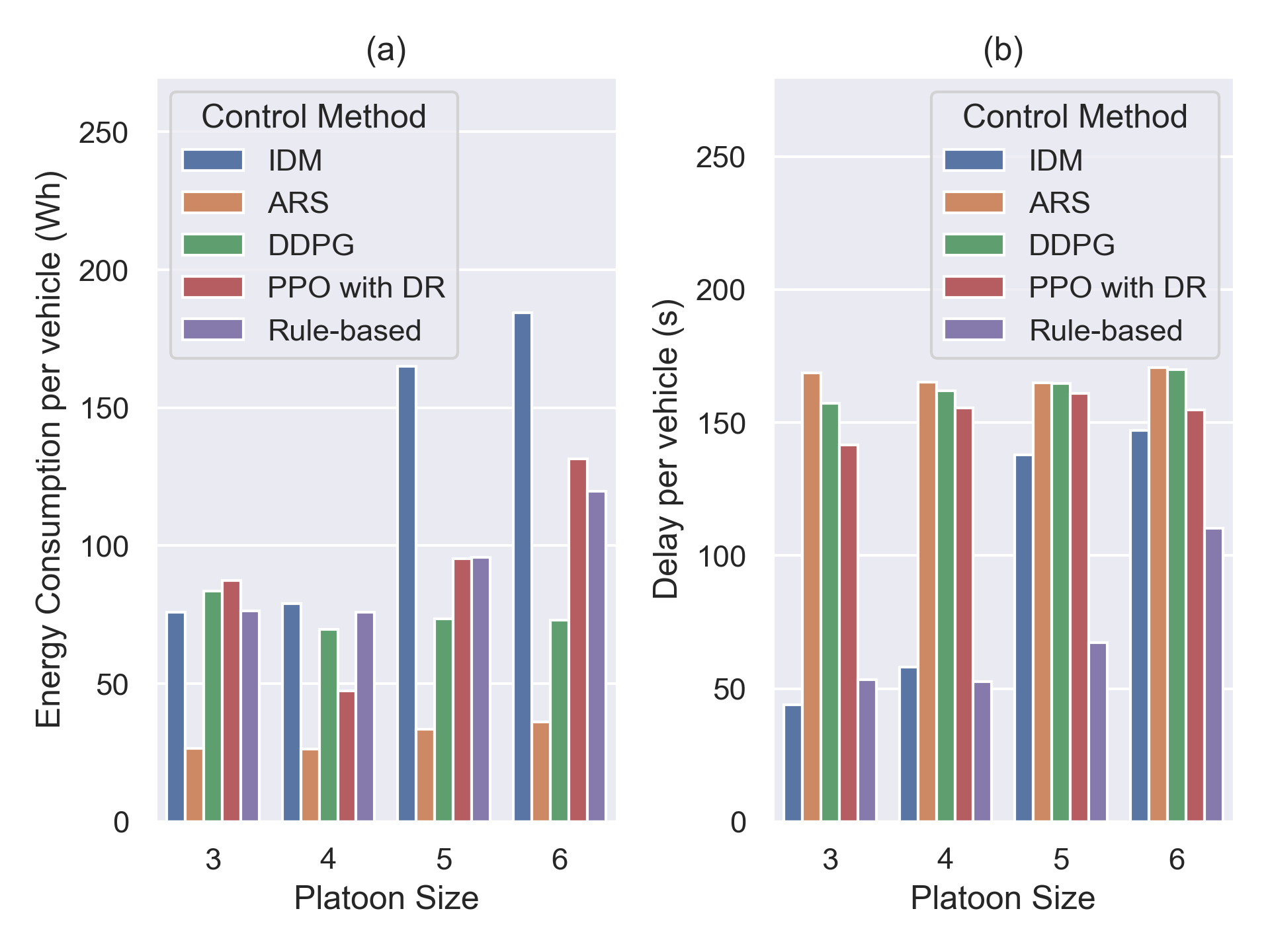}
    \caption{Single-vehicle-based Performance for Different Platoon Size. (a)Energy consumption. (b)Traffic delay.}
    \label{fig:fig8}
\end{figure}

\begin{figure*}[htb]
\begin{subfigure}{1\textwidth}
  \centering
  \includegraphics[width=1\linewidth]{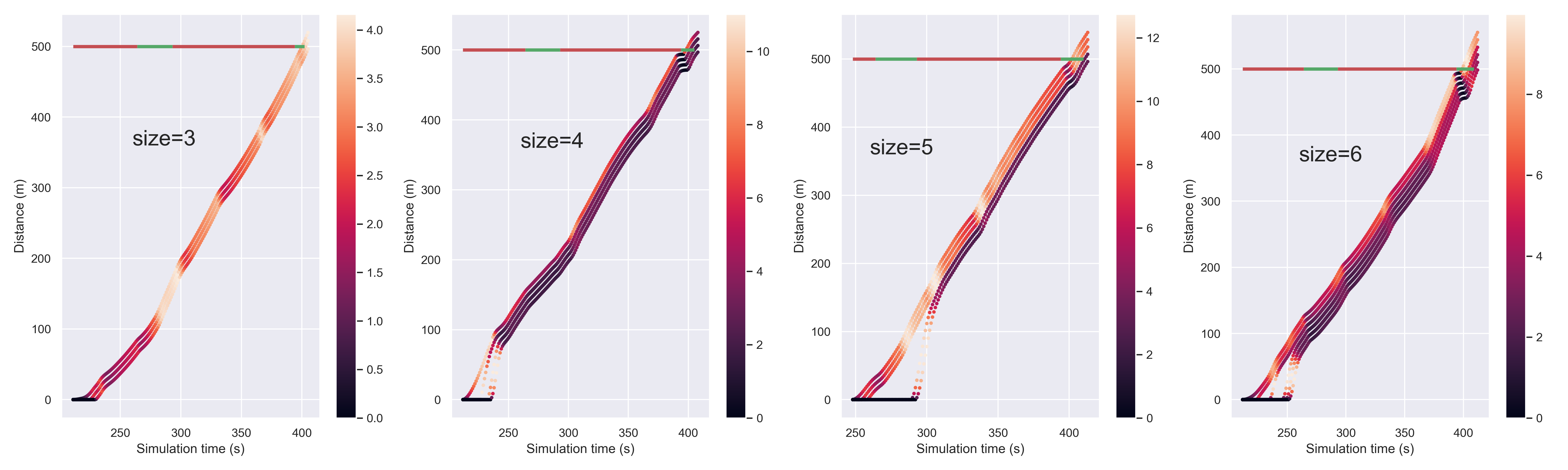}
  \subcaption{}
\end{subfigure}
\newline
\begin{subfigure}{1\textwidth}
  \centering
  \includegraphics[width=1\linewidth]{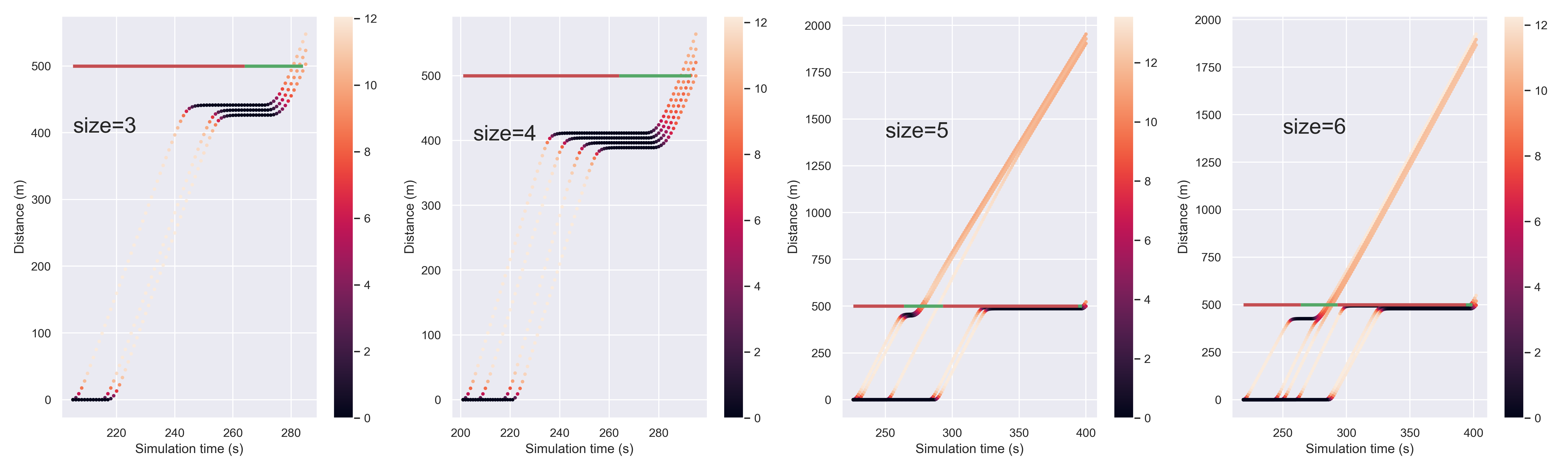}
  \subcaption{}
\end{subfigure}
\caption{The trajectories of the mixed platoon with different platoon size. (a) Controlled by ARS (b) Controlled by IDM}
\label{fig:fig9}
\end{figure*}

More specifically, we use the trained ARS agents with different platoon size configurations to run the simulations for evaluation. \red{We make comparison study with several other approaches and offer the single-vehicle-based statistical results for different platoon size configurations. Since optimization-based methods always suffer from unacceptable computation complexity, we only compare ARS with learning-based algorithms and rule-based algorithms: (1) IDM is regarded as paradigm of normal driving behavior, which resembles human drivers; (2) DDPG is deployed in several studies and achieves SOTA performance \cite{GUO2021102980, 8848852}; (3) PPO with DR setting serves as a baseline to observe the performance of DR settings; (4) Rule-based model \cite{trr2013}, which is known as Green Light Optimal Speed Advisory (GLOSA) system, provides CAVs with speed guidance in an "ego-efficient" way.}

The results are also collected from 10 independent simulations, which is illustrated in \autoref{fig:fig8}. It can be seen that the energy consumption and traffic delay increase sharply with the rise of platoon size for normal car-following approach (IDM), but the two indicators can maintain a stable level under ARS control. \red{As far as DDPG and PPO are concerned, the proposed ARS algorithm still achieves the optimal performance in terms of energy-related indicator, while the change of delay indicator is not significant. In addition, the consumed energy and time decline slightly when rule-based GLOSA system is employed, but this change is limited by its "ego-efficient" feature, which cannot takes the following HDVs into account. Although the traffic delay increases compared to IDM and GLOSA approach, it is just the result of extremely energy-saving setting due to the large ratio of weighting parameters $\omega_1$ and $\omega_2$, and we will show that the sacrificed travel delay can be reduced to approximately zero by regulating the parameters.}

For each platoon size configuration, the trajectories of the vehicles in the ego platoon are collected. We randomly sample several trajectories and draw the figures. The results are shown in \autoref{fig:fig9}. The color depth reflects the speed of the vehicles, while the horizontal line represents the phase of the traffic signal. Meanwhile, we implement an IDM-based study to make a comparison, and the sampled trajectories are also provided in \autoref{fig:fig9}. According to the figure, the ego platoon can cross the signalized intersection without any stops when the ego CAV is controlled by ARS. Thus, the unnecessary stop and rapid acceleration/deceleration can be avoided to promote energy conservation. In addition, the ego CAV can consider the crossing of more HDVs as the size of the platoon increases. When the number of HDVs exceeds 4, the platoon controlled purely by IDM can be divided so that some of the vehicles in the platoon cannot cross the intersection with the leading vehicles during the same phase. The ARS agents can adjust its velocity to relatively low value to fit the phase change and guarantee the effective operation of subsequent HDVs, while the CAVs controlled by IDM can only speed up if there is no interruption. This also illuminated that only based on the appropriate control methods can the comprehensive benefits of the CAVs be brought into traffic.

\subsection{The Impact of weighting parameters}
\begin{figure}[htb]
    \includegraphics[width=0.5\textwidth]{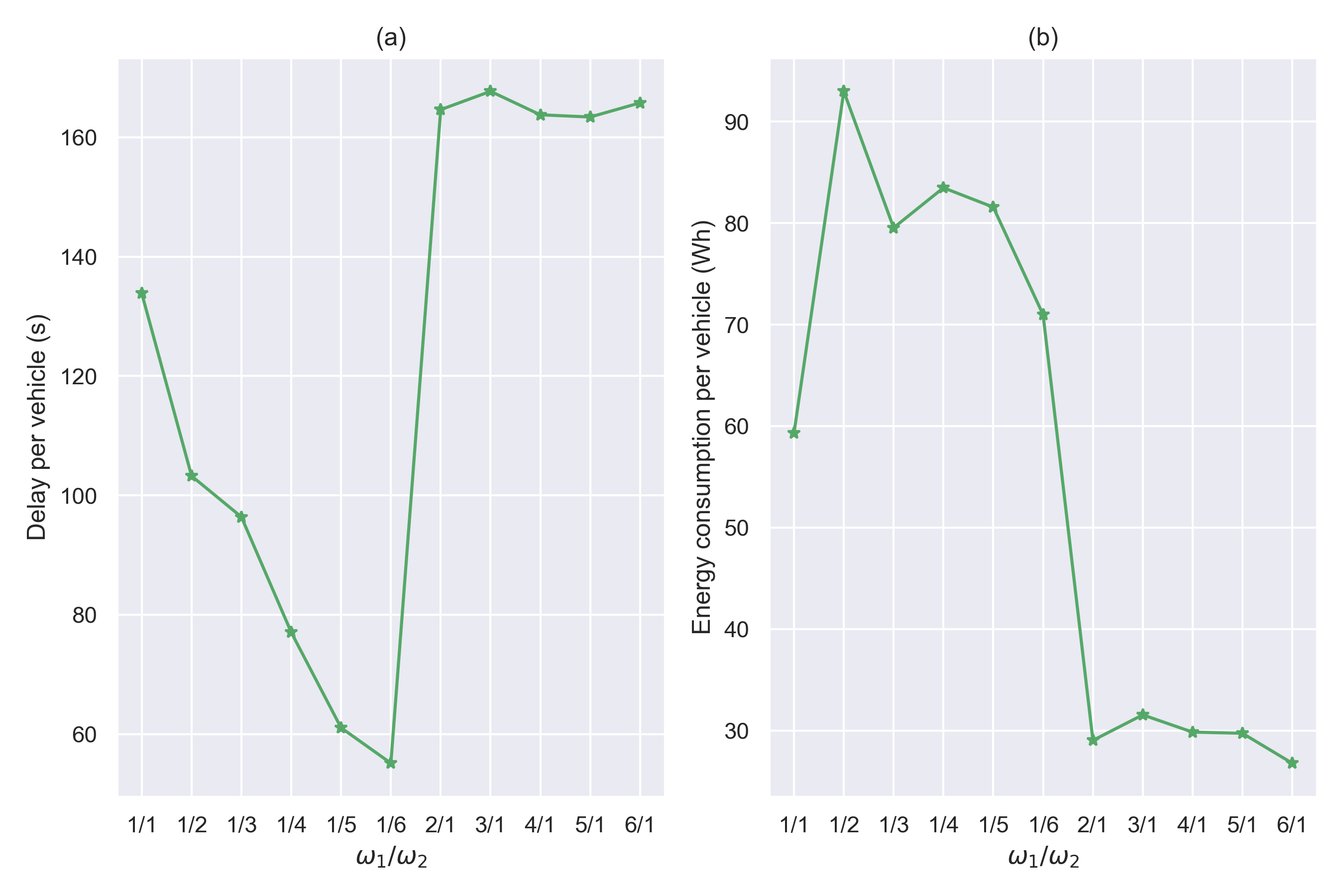}
    \caption{The variation curve with different settings of $\omega_1$ and $\omega_2$. (a) Delay per vehicle. (b) Energy consumption per vehicle.}
    \label{fig:fig10}
\end{figure}

The weighting parameters determine the optimization direction of the algorithm. Exploring the impact of the weighting parameters is valuable for understanding the effect of the ER-based reward signal. In particular, the impact of $\omega_1$ and $\omega_2$ mainly originates from the ratio (i.e., $\frac{\omega_1}{\omega_2}$) of the two values. Therefore, we fix $\omega_2$ to 1 and change $\omega_1$ from 1 to 6, and then fix $\omega_1$ to 1 with $\omega_2$ changing from 1 to 6. A "1+3" mixed platoon scenario is still taken as an example to observe the impact of the ratio, and the results are shown in \autoref{fig:fig10}. According to the figure, we can see that the delay of vehicles reduces rapidly with the increase of $\omega_2$ when $\omega_1 < \omega_2$. The policies learned by the agent can reduce both delay and energy consumption in these cases. When we set $\omega_1 > \omega_2$, the energy consumption can be reduced significantly. The policies in these cases can serve as economic driving strategies to maximize energy efficiency. 

\red{Simulations for other DRL algorithms with different weighting parameter settings are also carried out to make more comprehensive comparison studies, and the results are collected in \autoref{tab:sen}. It can be found that the performance of the proposed ARS-based control varies regularly with the change of weighting parameters, while the same outcome cannot be achieved by the other two DRL algorithms. This finding further enhances the flexibility and applicability of the framework with delay reward when considering regulating the relative importance between mobility and energy efficiency. Moreover, the ARS algorithm can achieve the optimal performance in terms of both travel delay and energy consumption. The significant decline of consumed electricity demonstrates that our method possess tremendous potential for the mixed platoon control task.}

\begin{table*}[htb]\caption{\red{Performance of the methods with different weighting parameter settings. Improvements for ARS compared with the IDM model are shown as Imp. The best performance is notated by bold style}}
\centering
\red{
\begin{tabular}{c|cccccccc}
\hline\hline
Ratio ($\frac{\omega_1}{\omega_2}$) & \multicolumn{3}{c}{Delay per vehicle (s)} & & \multicolumn{3}{c}{Energy consumption per vehicle (Wh)} &  \\
\Xcline{2-4}{0.4pt} \Xcline{6-8}{0.4pt}
  & PPO with DR & DDPG & ARS & Imp. $\uparrow$ & PPO with DR & DDPG & ARS & Imp. $\uparrow$ \\
\Xhline{1pt}
1/1 & 116.24 & 61.78 & 133.84 & -130.84\% & 99.47 & 89.78 & 59.30 & 61.27\% \\
1/2 & 107.44 & 162.51 & 103.24 & -78.07\% & 126.26 & 78.27 & 92.99 & 39.27\% \\
1/3 & 78.71 & 81.91 & 96.38 & -66.22\% & 106.04 & 106.77 & 79.54 & 48.05\% \\
1/4 & 96.77 & 81.98 & 77.04 & -32.88\% & 124.78 & 111.74 & 83.48 & 45.47\% \\
1/5 & 124.84 & 162.24 & 61.04 & -5.28\% & 106.10 & 77.70 & 81.58 & 46.72\% \\
1/6 & 98.64 & 161.71 & \textbf{55.11} & 4.95\% & 88.01 & 77.94 & 70.98 & 53.64\% \\
2/1 & 72.78 & 151.91 & 164.64 & -183.97\% & 113.87 & 77.81 & 29.03 & 81.04\% \\
3/1 & 67.78 & 79.64 & 167.71 & -189.26\% & 95.69 & 111.56 & 31.55 & 79.40\% \\
4/1 & 90.91 & 165.38 & 163.78 & -182.47\% & 112.64 & 77.79 & 29.84 & 80.51\% \\
5/1 & 98.91 & 159.24 & 163.41 & -181.83\% & 102.88 & 78.39 & 29.73 & 80.58\% \\
6/1 & 154.18 & 162.04 & 165.78 & -185.92\% & 48.40 & 69.68 & \textbf{26.79} & 82.51\% \\
\hline\hline
\end{tabular}
}
\label{tab:sen}
\end{table*}

More precisely, \red{compared with the basic IDM car-following behavior}, the electricity consumption is reduced by 39.27\% to 82.51\% with different weighting parameter settings. If we set $\omega_1=1$ with $\omega_2=6$, the energy can be saved by 53.64\% with approximately the same performance in terms of delay. \red{This result achieves SOTA performance when it is difficult to have both energy consumption and travel delay decline \cite{GUO2021102980, WEGENER2021102967}.}

\section{Conclusion}
In this paper, we propose a reinforcement learning framework to control the mixed platoon composed of CAVs and HDVs at a signalized intersection. By designing a novel state representation and reward function, the approach can be extended to the platoons with different platoon size. ARS is implemented to overcome the challenge caused by episodic reward, which is proved can outperform the distributed reward configuration for the utilized algorithm. Analysis and simulation results validate that ARS is capable of controlling the ego CAV and make the platoon cross the intersection without any stops. Meanwhile, great energy-efficient performance can be achieved, so we recommend the method as an economic driving strategy in practice. Being compared with several SOTA DRL algorithms, the proposed method gives a much higher reward with a simple architecture. 

It should be noted that the strategy put forward in this paper is still feasible with multi-intersection scenario by taking the SPaT information of the first downstream traffic signal as part of the state in succession. However, we only study the control of a single agent, while multi-agent cooperation may bring about more return. A collaboration way can be introduced with the support of vehicle-to-vehicle communication in this context.

As for the future research, firstly, the longitudinal motion of vehicles can be controlled by setting the acceleration in continuous action space. More comprehensive studies can start from the combination of longitudinal and lateral control in order to further tap the advantages of CAVs. \red{By designing proper strategy to incorporate car-following and lane-changing motion, the cooperative operation of CAVs from multi-lane traffic environment may has a profound influence on the overall mixed traffic performance}. Secondly, the influence of the traffic signal timing scheme is not explored in this paper, and it can be discussed specifically. Thirdly, the difference between traditional gasoline vehicles and electric vehicles can be discussed for the DRL-based adaptive control. \red{Finally, it is valuable to study the impact range of the ego CAV, which is determined by its sensing ability or communication ability, so as to make the model more practical.}. With the development of ITS, more reliable control methods will be implemented to create a sustainable and efficient urban traffic environment.



\ifCLASSOPTIONcaptionsoff
  \newpage
\fi



%
\bibliographystyle{IEEEtran}
\bibliography{IEEEexample, IEEEabrv}

%


\end{document}